\documentstyle[11pt,newpasp,twoside,epsf]{article}
\def\eps@scaling{.95}
\def\epsscale#1{\gdef\eps@scaling{#1}}

\def\plotone#1{\centering \leavevmode
    \epsfxsize=\eps@scaling\columnwidth \epsfbox{#1}}

\def\edcomment#1{\iffalse\marginpar{\raggedright\sl#1\/}\else\relax\fi}
\reversemarginpar

\begin{document}
\title{Spectroscopic Diagnostics for AGNs}
 \author{Sylvain Veilleux}
\affil{Department of Astronomy, University of Maryland, College Park, MD 20742}

\begin{abstract}
A review of the spectroscopic tools needed to characterize AGNs is
presented. This review focusses on ultraviolet, optical and infrared
emission-line diagnostics specifically designed to help differentiate
AGNs from starburst-dominated galaxies. The strengths and weaknesses
of these methods are discussed in the context of on-going and future
AGN surveys.
\end{abstract}

\section{Introduction}

The first decade of the 21st Century promises to become the Golden Age
of extragalactic astronomy. The 2dF and Sloan Digital Sky Surveys have
already made significant contributions to our knowledge of the
extragalactic universe. On-going and planned wide-field (pencil-beam)
imaging and spectroscopic surveys with 4m (8m)-class telescopes from
the ground and in space (e.g., SIRTF, SOFIA, Herschel, NGST) will
nicely complement these large-scale surveys and should go a long way
to answer some of the most fundamental questions in extragalactic
astronomy: How do galaxies form? How do they evolve? How do
supermassive black holes fit in this picture of galaxy formation?
Which objects are the main contributors to the overall energy budget
of the universe? To properly answer these questions, one will need to
differentiate objects powered by nuclear fusion in stars (i.e.~normal
and starburst galaxies) from objects powered by mass accretion onto
supermassive black holes (quasars and AGNs). A wide variety of
diagnostic tools have been used in the past for this purpose with
different degree of success.

Due to space limitations, the present discussion focusses on {\em
emission-line} diagnostics.  The fundamental principles behind these
diagnostics are reviewed in \S 2. Next, the main diagnostic tools
available in the ultraviolet, optical, and infrared domains are
described in \S 3, \S 4, and \S 5, respectively. A table listing the
main diagnostic lines is given in each of these sections. Additional
factors which may complicate the use of these tools are discussed in
\S 6. A summary is given in \S 7 along with an outlook on the
future. Note that this review is not meant to be exhaustive; it is
meant to emphasize the practical aspects of starburst/AGN spectral
classification.  Readers who are looking for a more detailed
discussion of the physics behind these diagnostic tools should refer
to the original papers listed in the text.

\section{Basic Principles}

Activity driven by mass accretion onto supermassive black holes
differs in many ways from star-formation activity.  The thermal and
non-thermal processes associated with the accretion disk and its
surroundings (e.g., corona) are at the origin of the ``hard'' ionizing
continuum detected in quasars and AGNs (e.g., Krolik 1999). Material
in the vicinity of the nucleus will bear the imprint of this strong
radiation field.  The deep gravitational potential at the center of
these galaxies allows the presence of high-density ($\ga$ 10$^9$
cm$^{-3}$), high-velocity ($\ga$ 2000 km s$^{-1}$) gas clouds in the
inner parsec of quasars and AGNs. This so-called broad-line region or
BLR is a powerful diagnostic of nuclear activity in galaxies.  The
main signatures of the BLRs are broad recombination lines which are
unaffected by the effects of collisional de-excitation at high
densities. Two general methods have been used in the past to detect
BLRs in galaxies: direct spectroscopy and spectropolarimetry. This
last method relies on the presence of dust or electrons (``mirrors'')
to scatter the BLR signature towards the line of sight (e.g.,
Antonucci 1993). Direct spectroscopy searches for the presence of the
broad recombination lines at wavelengths where the effects of dust
extinction are reduced. As shown in Table 1 for representative
Galactic extinction (see, e.g., Cardelli, Clayton, \& Mathis 1989;
Draine \& Lee 1984; Draine 1989; Lutz et al. 1996; Lutz 1999), great
increase in sensitivity can in principle be obtained by observing at
longer wavelengths.

\begin{table*}[h]
\caption{Galactic Dust Extinction and Column Densities}
\footnotesize
\begin{center}
\begin{tabular}{lll}
\tableline
\tableline
\noalign{\vskip 7.5 pt}
$\lambda$ &$\tau(\lambda)/\tau({\rm H}\alpha)$ & N$_{\rm H}$(cm$^{-2}$) @ $\tau(\lambda)$ = 1\\
\tableline
\noalign{\vskip 7.5 pt}
Ly$\alpha$ 1216 \AA& 2.0 - 4.5  & 0.5 -- 1.0 $\times$ 10$^{21}$\\
V band 5500 \AA& 1.2 & 1.7 $\times$ 10$^{21}$\\
H$\alpha$ 6563 \AA& 1.0 & 2.2 $\times$ 10$^{21}$\\
J band 1.25 $\mu$m&1/3 & 6.1 $\times$ 10$^{21}$\\
H band 1.65 $\mu$m&1/4.5 & 9.8 $\times$ 10$^{21}$\\
K band 2.2 $\mu$m& 1/7 & 1.6 $\times$ 10$^{22}$\\
L band 3.4 $\mu$m& 1/15 & 3.4 $\times$ 10$^{22}$\\
M band 5.0 $\mu$m& 1/30 & 6.4 $\times$ 10$^{22}$\\
N band 10 $\mu$m& 1/15 & 3.2 $\times$ 10$^{22}$\\
12 $\mu$m &1/30 & 6.2 $\times$ 10$^{22}$ \\
25 $\mu$m &1/60 &1.3 $\times$ 10$^{23}$\\
60 $\mu$m & 1/400 &8.6 $\times$ 10$^{23}$\\
100 $\mu$m & 1/700 & 1.5 $\times$ 10$^{24}$\\
\noalign{\vskip 7.5 pt}
\tableline
\end{tabular}
\end{center}
\end{table*}

In highly obscured objects with N$_{\rm H}$ $\ga$ 10$^{24}$ cm$^{-2}$,
direct detection of the BLRs becomes very difficult and one has to
rely on spectropolarimetry to search for the presence of a BLR.  The
obscuring screen may not be opaque in all directions, however. The
ionizing radiation field may be able to escape in certain directions
and ionize the surrounding material on scales beyond the obscuring
material. Distributed in the shallower portion of the gravitational
potential ($\sim$ 0.1 -- 1 kpc), this ``narrow-line region'' or NLR is
another excellent probe of nuclear activity. The ionizing spectra of
all but the hottest O stars cut off near the He II edge (54.4 eV;
Dopita et al. 1995). In contrast, the ionizing spectrum of AGNs
contains a relatively large fraction of high-energy photons (e.g.,
Elvis et al. 1994).  Optically thick gas clouds ionized by the hard
continuum of AGNs will present a stratified ionization structure with
(1) a highly ionized inner face (closest to the AGN), (2) a large
partially zone with characteristic fraction of ionized hydrogen
H$^+$/H $\sim$ 0.2 -- 0.4 produced by the deposition of keV X-rays
(recall that the absorption cross sections of H$^0$, He$^0$, and all
other ions decrease rapidly with increasing energy; Osterbrock 1989),
and (3) a neutral zone facing away from the AGN. The fast free
electrons in the partly ionized zone will have a positive effect on
the strengths of low-ionization lines produced by collisional effects,
while the highly ionized conditions in the inner face will favor the
production of emission lines from ions with high ionization potentials
(e.g., Ferland \& Netzer 1983; Ferland \& Osterbrock 1986, 1987;
Binette, Wilson, \& Storchi-Bergmann 1996).

Based on these physical principles, one should choose narrow emission
line diagnostics following ten basic rules or ``Commandments'' (a
reminder of the 1700th anniversary of the adoption of Christianity as
a national religion in Armenia):

\begin{itemize}
\item[1.] Thou shalt use lines which emphasize the differences between
H~II regions and AGNs; i.e., use high-ionization lines or low-ionization
lines produced in the partially ionized zone.
\item[2.] Thou shalt use strong lines which are easy to measure in
typical spectra.
\item[3.] Thou shalt avoid lines which are badly blended with other
emission or absorption line features.
\item[4.] Thou shalt use lines with small wavelength separation to
minimize sensitivity to reddening.
\item[5.] Thou shalt use line ratios from the same elements or involving
hydrogen recombination lines to eliminate or reduce abundance
dependence.
\item[6.] Thou shalt avoid lines from Mg, Si, Ca, Fe -- depleted onto
dust grains.
\item[7.] Thou shalt use lines easily accessible to current
UV/optical/IR detectors.
\item[8.] Thou shalt avoid lines affected by strong stellar absorption
features.
\item[9.] Thou shalt avoid lines affected by strong atmospheric features.
\item[10.] Thou shalt use lines at long wavelengths to reduce the effects
of dust extinction.
\end{itemize}

\section{Ultraviolet Emission-Line Diagnostics}

When possible, ultraviolet diagnostic tools should be avoided because
of their sensitivity to dust extinctions (see the Tenth Commandment
and Table 1).  However, investigators of the high-redshift universe
often have very little choice but to study this region of the
electromagnetic spectrum.  The ultraviolet domain is potentially a
rich source of diagnostic lines.  The main emission lines are listed
in Table 2. Among the most useful diagnostics to discriminate between
AGNs and starbursts are the N~V $\lambda$1240/He~II $\lambda$4686, N~V
$\lambda$1240/C~IV $\lambda$1549, N~V $\lambda$1240/Ly$\alpha$, and
C~IV $\lambda$1549/Ly$\alpha$ emission-line ratios. As shown in Figure
1, these ratios are sensitive functions of the shape of the ionizing
continuum (harder spectra provide more heating per photoionization,
therefore increasing the temperature). These line ratios have been
used extensively in studies of high-$z$ quasars (e.g., Hamann \&
Ferland 1999) and radio galaxies (e.g., R\"ottgering et al. 1997;
Villar-Martin et al. 1996, 1999), and the analysis of low-$z$
AGNs/LINERs (e.g., Ho et al. 1996; Barth et al. 1996, 1997; Maoz et
al. 1998; Nicholson et al. 1998) and starburst galaxies (e.g., Robert,
Leitherer, \& Heckman 1993).
 
\begin{table*}[h]
\footnotesize
\caption{Ultraviolet Emission-Line Diagnostics}
\begin{center}
\begin{tabular}{llll}
\tableline
\tableline
\noalign{\vskip 7.5 pt}
\multicolumn{2}{c}{Low-to-Moderate Ionization Lines} & \multicolumn{2}{c}{High-Ionization Lines}\\
\noalign{\vskip 7.5 pt}
Line & $\chi$(eV) & Line & $\chi$(eV)\\
\tableline
\noalign{\vskip 7.5 pt}
C III 977 \AA & 24.4 & O VI 1032, 1038 \AA & 114\\
N III 991, 1750 \AA & 29.6 & N V 1240 \AA & 77.4\\
Ly$\beta$ 1026 \AA, Ly$\alpha$ 1216 \AA & 13.6 &  O IV] 1407 \AA & 54.9 \\
Si IV 1394, 1403 \AA & 33.5 & N IV] 1488 \AA & 47.4 \\  
O III] 1663 \AA & 35.1 &C IV 1549 \AA & 47.9 \\    
N III] 1750 \AA & 29.6 &He II 1085, 1640 \AA & 54.4 \\ 
Si III 1895 \AA & 16.3\\
C III] 1909 \AA & 24.4\\
Fe II 2080, 2500, 3300 \AA & 7.9\\
$[$O III] 2322 \AA & 35.1 \\
C II] 2326 \AA & 11.3\\
Si II 2336 \AA & 8.2\\
Mg II 2798 \AA & 7.6\\
\noalign{\vskip 7.5 pt}
\tableline
\end{tabular}
\end{center}
\end{table*}

\begin{figure}
\plotone{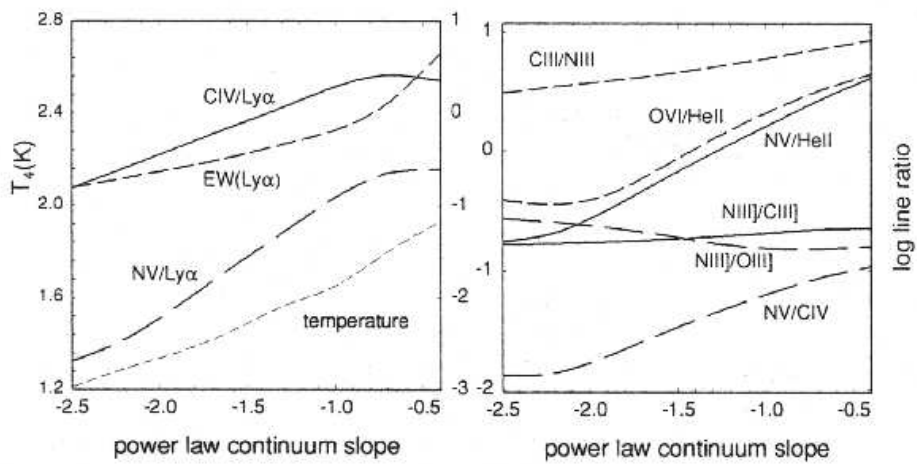}
\caption{Predicted UV line flux ratios, gas temperatures, and
dimensionless equivalent widths in Ly$\alpha$ for clouds photoionized
by different power-law spectra ($\nu^\alpha$). The UV-to-X-ray slopes
of QSOs are roughly consistent with $\alpha \approx -1.5$. From Hamann
\& Ferland (1999). }
\end{figure}

\section{Optical Emission-Line Diagnostics}

The excellent quantum efficiency of current CCDs combined with the
high transparency and low emissivity of the Earth's atmosphere at
optical wavelengths make optical spectroscopy the easiest way to
identify AGNs.  Table 3 lists the strongest diagnostic lines between
3000 \AA\ and 1 $\mu$m. Classification schemes involving several line
ratios which take full advantage of the physical distinction between
the two types of objects and minimize the effects of reddening
correction and errors in the flux calibration have proven very useful
for the identification of galaxies as AGNs or starbursts (e.g.,
Phillips, Baldwin, \& Terlevich 1981; Veilleux \& Osterbrock 1987;
Osterbrock, Tran, \& Veilleux 1992; Dopita et al. 2000). Examples of
emission-line diagrams are shown in Figure 2 for ultraluminous
infrared galaxies from the 1-Jy sample (ULIGs; these are IRAS galaxies
with infrared luminosities between 8 and 1000 $\mu$m larger than or
equal to 10$^{12}$ L$_\odot$; Kim 1995).  The results from this classification
indicates that the fraction of Seyfert nuclei increases from $\sim$
5\% at log[L$_{\rm IR}$/L$_\odot$] = 10 -- 11, to $\sim$ 50\% at
log[L$_{\rm IR}$/L$_\odot$] $>$ 12.3 (Veilleux et al. 1995, Kim,
Veilleux, \& Sanders 1998; Veilleux, Kim, \& Sanders 1999a).
\begin{table*}[h]
\footnotesize
\caption{Optical Emission-Line Diagnostics}
\begin{center}
\begin{tabular}{llll}
\tableline
\tableline
\noalign{\vskip 7.5 pt}
\multicolumn{2}{c}{Low-to-Moderate Ionization Lines} & \multicolumn{2}{c}{High-Ionization Lines}\\
\noalign{\vskip 7.5 pt}
Line & $\chi$(eV) & Line & $\chi$(eV)\\
\tableline
\noalign{\vskip 7.5 pt}
$[$O II] 3727, 7325 \AA & 13.6 & $[$Ne V] 3346, 3426 \AA & 97.1\\
$[$Ne III] 3869, 3968 \AA & 41.0 & $[$Fe V] 3840, 3893, 4071 \AA & 54.8\\
$[$O III] 4363, 5007 \AA & 35.1 & $[$Fe VII] 3588, 3760, 4071, 5721, 6087 \AA & 99.0\\
Fe II 4500, 5190, 5300 \AA & 7.9 & He II 4686 \AA & 54.4 \\
H$\beta$ 4861 \AA, H$\alpha$ 6563 \AA & 13.6 & $[$Fe XIV] 5303 \AA & 344 \\
He I 5876, 7065 \AA & 24.6& $[$Fe X] 6375 \AA & 235\\
$[$O I] 6300, 6363 \AA & 0.0 & $[$Fe XI] 7892 \AA & 262 \\
$[$N II] 5755, 6548, 6583 \AA & 14.5 \\
$[$S II] 6716, 6731 \AA & 10.4\\
$[$S III] 6312, 9069, 9531 \AA & 23.3\\
\noalign{\vskip 7.5 pt}
\tableline
\end{tabular}
\end{center}
\end{table*}

\begin{figure}[ht]
\epsscale{0.5}
\plotone{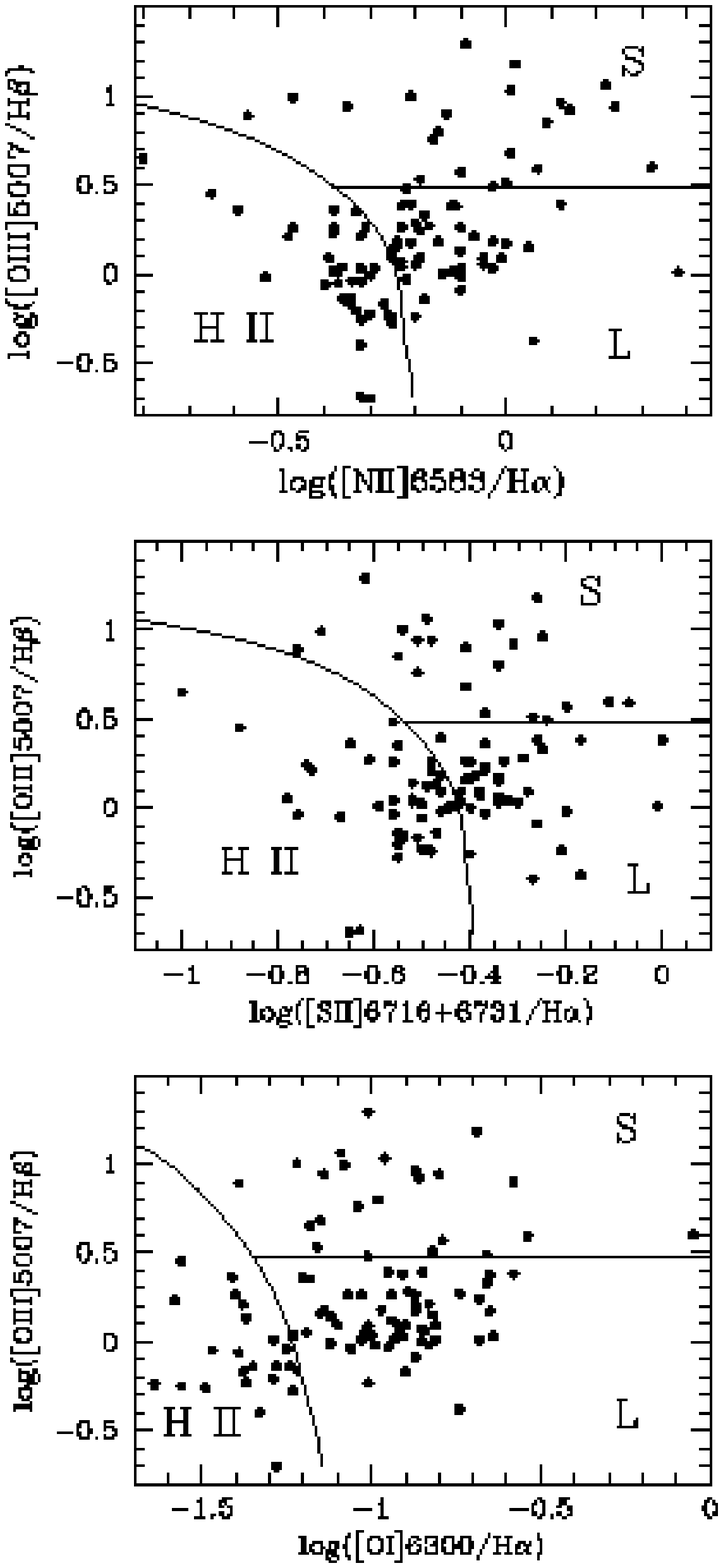}
\caption{Examples of optical line ratio diagrams used for the
classification of ultraluminous infrared galaxies. From Veilleux et
al. (1999a).}
\end{figure}

\section{Near-Infrared Emission-Line Diagnostics}

Infrared-bright galaxies such as those discussed in the previous
section are hosts to large quantities of molecular gas and dust (e.g.,
Solomon et al. 1997). The optical line ratios measured in these
objects are undoubtedly affected by dust extinction. It is therefore
important to also observe these objects at longer wavelengths to verify the
results derived from the optical spectra. Near-infrared spectroscopy
has had success finding obscured BLRs in several ULIGs
(e.g., Hines 1991; Veilleux et al. 1997b, 1999b;
spectropolarimetry has lent support to some of these findings: Hines
1991; Hough et al. 1991; Hines \& Wills 1993; Hines et al. 1995; Young
et al. 1993). This technique has also proven useful in the study of
highly reddened BLRs in intermediate Seyferts (1.8's and 1.9's;
Goodrich 1990; Rix et al. 1990) and in optically classified Seyfert~2
and radio galaxies (e.g., Blanco, Ward, \& Wright 1990; Goodrich,
Veilleux, \& Hill 1994; Ruiz, Rieke, \& Schmidt 1994; Hill, Goodrich,
DePoy 1996; Veilleux, Goodrich, \& Hill 1997a).

The line of choice for ground-based near-infrared searches of obscured
BLRs in nearby galaxies is Pa$\alpha$ at 1.8751 $\mu$m (Table 4).
Under Case B recombination (Osterbrock 1989), this line is one-third
the strength of H$\alpha$ and is {\it twelve} times stronger than
Br$\gamma$~$\lambda$2.1655, the next best diagnostic line (e.g.,
Goldader et al. 1995). This huge gain in intensity more than
compensates the slightly larger optical depth due to extinction at the
shorter wavelength of Pa$\alpha$ (see 4th column in Table 4).

Another important AGN diagnostic line in the K band is
[Si~VI]~$\lambda$1.962.  The existence of five-times ionized silicon
ions requires energies larger than 167 eV (Table 4). This forbidden
line has been detected in a number of optically selected Seyfert~2
galaxies with a strength comparable to that of [Fe~VII]~$\lambda$6087
(roughly a tenth the strength of H$\beta$), as expected from
photoionization by a AGN power-law continuum (Oliva \& Moorwood 1990;
Greenhouse et al. 1993; Marconi et al. 1994; Oliva et al. 1994;
Thompson 1995, 1996).  Near-infrared spectroscopic surveys of ULIGs
have confirmed the optical results: the fraction of objects with
genuine AGNs (with a BLR or strong [Si~VI] $\lambda$1.962 feature) is
at least $\sim$ 20 -- 25\%, but reaches $\sim$ 35 -- 50\% for those
objects with log[L$_{\rm ir}$/L$_\odot$] $>$ 12.3.  Nevertheless, the
presence of an AGN in ULIGs does not necessarily imply that AGN
activity is the dominant source of energy in these objects. A more
detailed look at the AGNs in these ULIGs is needed to answer this
question.

\begin{table*}[h]
\footnotesize
\caption{Hydrogen Recombination Lines and some High Ionization Lines 
in the Near-Infrared}
\begin{center}
\begin{tabular}{llllll}
\tableline
\tableline
\noalign{\vskip 7.5 pt}
\multicolumn{4}{c}{Hydrogen Recombination Lines} & \multicolumn{2}{c}{High-Ionization Lines}
\\Line & $\lambda$($\mu$m) & F/F$_{{\rm H}\alpha}$& A$_\lambda$/A$_{{\rm H}\alpha}$ & Line & $\chi$(eV)\\
\tableline
\noalign{\vskip 7.5 pt}
H$\beta$ & 0.4861 & 1.00 & 1.48    & $[$S IX] 1.252 $\mu$m & 328\\   
H$\alpha$ & 0.6563 & 2.85 & 1.00   & $[$Si X] 1.430 $\mu$m & 351\\   
Pa$\gamma$ & 1.0938 & 0.090 & 0.45 & $[$Si XI] 1.932 $\mu$m & 401\\  
Pa$\beta$ & 1.2818 & 0.162 & 0.34  & $[$Si VI] 1.962 $\mu$m & 167\\  
Pa$\alpha$ & 1.8751 & 0.332 & 0.18 & $[$Ca VIII] 2.321 $\mu$m & 128\\
Br$\gamma$ & 2.1655 & 0.0275 & 0.14& $[$Si VII] 2.483 $\mu$m & 205\\ 
Br$\alpha$ & 4.0512 & 0.0779 & 0.05& $[$Si IX] 3.935 $\mu$m & 303\\  
\noalign{\vskip 7.5 pt}
\tableline
\end{tabular}
\end{center}
\end{table*}

A strong linear correlation has long been known to exist between the
continuum (or, equivalently, bolometric) luminosities of broad-line
AGN and their emission-line luminosities (e.g., Yee 1980; Shuder 1981;
Osterbrock 1989).  This correlation has often been used to argue that
the broad-line regions in AGNs are photoionized by the nuclear
continuum. If this is the case, the broad-line--to--bolometric
luminosity ratio is a measure of the covering factor of the BLR (e.g.,
Osterbrock 1989). This correlation can be used to estimate the
importance of the AGN in powering ultraluminous infrared galaxies
(Veilleux et al. 1997b, 1999b). In ULIGs powered uniquely by an AGN,
we expect the broad-line luminosities to fall along the correlation
for AGNs. Any contribution from a starburst will increase the
bolometric luminosity of the ULIG without a corresponding increase in
the broad-line luminosity.  Starburst-dominated ULIGs are therefore
expected to fall below the ``pure-AGN'' correlation traced by the
optical quasars in a diagram of $L_{\rm H\beta}$(BLR) plotted as a
function of $L_{\rm bol}$. The data of ULIGs with optical and obscured
BLRs are shown in Figure 3.  A discussion of the methods and
assumptions which were used to create this figure is presented in
Veilleux et al. (1999a). Figure 3 strongly suggests that most ($\sim$
80\%) of the ULIGs with optical or near-infrared BLRs in the 1-Jy
sample are powered predominantly by the quasar rather than by a
powerful starburst. In other words, {\em the detection of an optical
or near-infrared BLR in a ULIG (about 20\% of the total 1-Jy sample)
appears to be an excellent sign that the AGN is the dominant energy
source in that ULIG.}

\begin{figure}
\epsscale{0.7}
\plotone{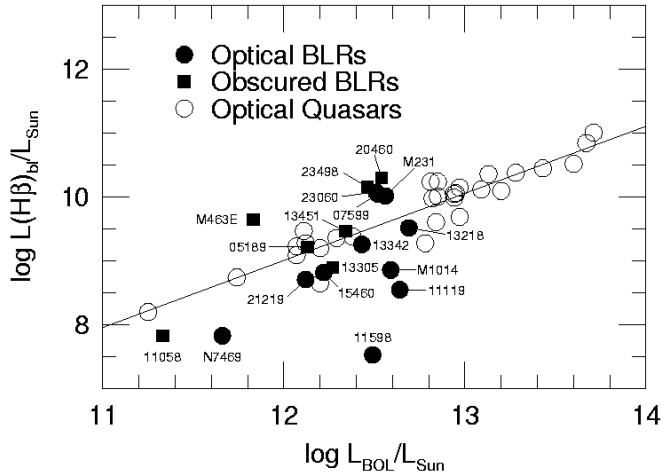}
\caption{Dominant energy source of ultraluminous infrared galaxies
based on their broad-line luminosities. The solid line is the best fit
for the optical quasars. From Veilleux et al. (1999b).}
\end{figure}

\section{Mid-to-Far Infrared Emission-Line Diagnostics}

The mid-to-far infrared region has long been known to be a rich source
of emission-line diagnostics in active and starburst galaxies (e.g.,
Watson et al. 1984; Roche et al. 1984; Aitken \& Roche 1985; Crawford
et al. 1985; Lugten et al. 1986; Duffy et al. 1987; Roche et al. 1984,
1991, Spinoglio \& Malkan 1992; Voit 1992). With the Infrared Space
Observatory (ISO) much progress has been made in this area of research
in recent years (see review by Genzel \& Cesarsky 2000). One of the
most important applications of ISO spectroscopy has been its use as a
tool to distinguish between star formation and AGN activity in
obscured environments. High ionization fine structure lines are strong
in the NLR of AGNs but very weak in starbursts (Table
5). Fine-structure lines have smaller excitation energies than their
optical counterparts, so they are less temperature sensitive
and less model dependent.  Unfortunately, they are also fainter than
their optical counterparts, and are therefore difficult to detect even
in genuine, optically-selected AGNs (e.g., Genzel et al. 1998).

\begin{table*}[h]
\footnotesize
\caption{Strongest Fine-Structure Lines expected from AGNs}
\begin{center}
\begin{tabular}{ll}
\tableline
\tableline
\noalign{\vskip 7.5 pt}
Line & $\chi$(eV)\\
\tableline
\noalign{\vskip 7.5 pt}
$[$Ar III] 9 $\mu$m & 27.6\\
$[$S IV] 10.5 $\mu$m & 34.8\\
$[$Ne II] 12.8 $\mu$m & 21.6\\
$[$Ne V] 14.3, 24.2 $\mu$m & 97.1\\
$[$Ne III] 15.6, 36.0 $\mu$m & 41.0\\
$[$S III] 18, 34 $\mu$m & 23.3\\
$[$O IV] 26 $\mu$m & 54.9\\
$[$Si II] 35 $\mu$m & 8.2\\
$[$O III] 52, 88 $\mu$m & 35.1\\
\noalign{\vskip 7.5 pt}
\tableline
\end{tabular}
\end{center}
\end{table*}

This work has since been extended to larger samples, using the
polycyclic aromatic hydrocarbon (PAH) diagnostic to reach fainter
sources (e.g., Lutz et al. 1998; Rigopoulou et al. 1999; Tran et
al. 2001). The PAH features at 3.3, 6.2, 7.7, 8.7, and 11.2 $\mu$m are
ubiquitous in normal galaxies and starbursts but absent near an AGN
(e.g., Roche et al. 1991). Obscured regions also show absorption
features, the strongest ones at 9.7 and 18 $\mu$m being due to
silicate dust, which complicate the placement of the continuum near
the PAH features.  An object-by-object comparison of the optical and
ISO spectral types for ULIGs in the 1-Jy sample reveals a remarkably
good agreement between the two classification schemes if optically
classified LINERs are assigned to the starburst group (Fig. 4). These
results indicate that strong AGN activity, once triggered, quickly
breaks the obscuring screen at least in certain directions, thus
becoming detectable over a wide wavelength range.

\begin{figure}
\epsscale{0.35}
\plotone{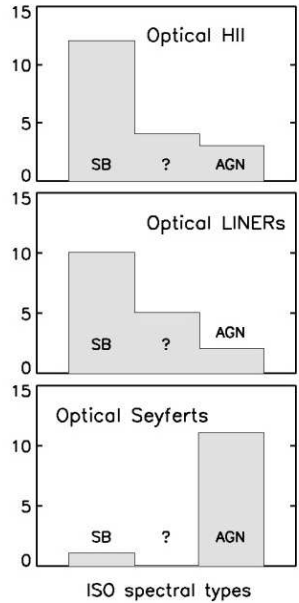}
\caption{Optical versus mid-infrared classification of ultraluminous
infrared galaxies. From Lutz, Veilleux, \& Genzel (1999).}
\end{figure}

\section{Complications}

\subsection{Contribution from Shocks}

Violent gas motions associated with AGN-driven or starburst-driven
outflows or galaxy mergers may cause shock waves with velocities of
100 -- 500 km s$^{-1}$ in the ISM of the host galaxies.  The shocks
may produce a strong flux of EUV and soft X-ray radiation which may be
absorbed in the shock precursor H~II region (e.g., Sutherland,
Bicknell, \& Dopita 1995). The combination of the low-ionization
emission-line spectrum from the post-shock material and the
high-ionization emission-line spectrum from the precursor H~II region
can reproduce many of the spectroscopic signatures of LINERs and
narrow-line AGNs (e.g., Dopita \& Sutherland 1995). Fortunately, there
are important physical differences between shock ionization and
photoionization by an AGN (e.g., Morse, Raymond, \& Wilson
1996). First, the line ratios produced in photoionized objects should
be independent of the gas kinematics, while they are expected to
correlate with the kinematics of the shock-ionized material. This
effect is seen in a few optically and infrared-selected LINERs
(Veilleux et al. 1994, 1995).  The ionizing ultraviolet and soft X-ray
continuum in shock-ionized objects should be extended on the same
scale as the shock structure, while it is expected to be a point
source in the case of pure AGN photoionization. Finally, the electron
temperature in shock ionized objects is expected to be considerably
higher. Temperature-sensitive line ratios such as C~III
$\lambda$1909/$\lambda$977 and N~III $\lambda$1750/$\lambda$991 in the
ultraviolet and [O~III] $\lambda$5007/$\lambda$4363 and [N~II]
$\lambda$5755/$\lambda$6583 in the optical range are the prime
diagnostics of shock excitation. This method was used by Kriss et
al. (1992) to deduce that shock excitation is likely to be important
in the NLR of NGC~1068.

\subsection{Aperture Effects}

Circumnuclear starbursts often accompany AGNs (see, e.g., recent
reviews by Veilleux 2000 and Gonzales Delgado 2001). The strength of
the AGN signature is therefore a function of the size of the
extraction aperture. This effect is particularly evident among
infrared-selected galaxies where circumnuclear starbursts are nearly
always present. Figure 5 shows the line ratios of luminous infrared
galaxies as function of aperture size. The line ratios in some of
these objects are seen to drift towards the H~II region locus with
increasing aperture size; large apertures dilute the AGN
signature. Aperture effects will be particularly important in samples
which cover a broad redshift range where a constant angular aperture
corresponds to a wide range in linear scale. For a meaningful
statistical analysis of the spectral classification one should use a
fixed {\em linear} aperture for all objects in the sample (regardless
of redshifts).

\begin{figure}
\epsscale{0.75}
\plotone{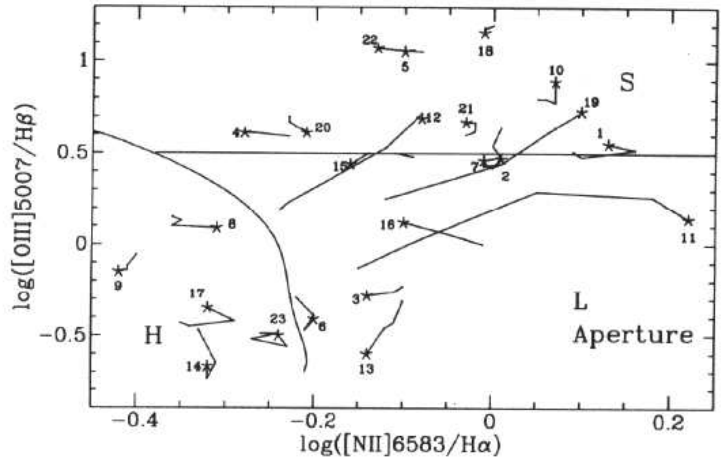}
\caption{Aperture effects. Line ratios as a function of the size of
the extraction aperture. The asterisks mark the nuclear values. The
size of the extraction aperture increases (generally doubles) between
each data point. From Veilleux et al. (1995).}
\end{figure}

\subsection{Morphological Biases}

Strong trends exist between the presence of an AGN, the mid-to-far
infrared colors, and the host morphology. Objects with ``warm''
infrared colors (e.g., IRAS $f_{25}/f_{60} >$ 0.2) often harbor an AGN
at optical or near-infrared wavelengths or in polarized light (de
Grijp et al. 1985; Veilleux et al. 1995, 1997b, 1999ab; Heisler,
Lumsden, \& Bailey 1997).  Infrared-selected samples are often biased
towards or against the presence of AGNs (but this is not the case for
the 1-Jy sample; Kim \& Sanders 1998). The same thing can be said
about galaxy morphology.  ULIGs often show signs of galaxy
interactions. Most ULIGs are involved in the merger of two relatively
large galaxies. Optically-classified Seyferts (especially those of
type 1) are generally found in advanced mergers, while H~II galaxies
and LINERs are found in all merger phases (Fig. 6; see also Veilleux
2001).  This means that surveys which specifically look for compact
objects will be biased against starburst galaxies and are not
statistically reliable for spectral classification purposes.

\begin{figure}
\epsscale{0.75}
\plotone{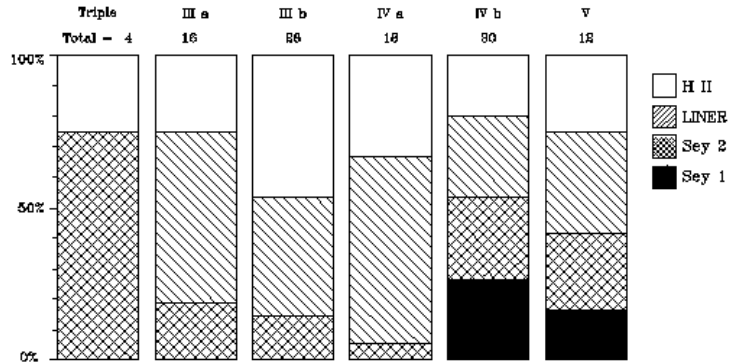}
\caption{Morphological biases among infrared-selected AGNs. The hosts
of ultraluminous infrared galaxies which are optically classified as
Seyferts generally are advanced mergers (morphological classes IVa,
IVb, or V). From Veilleux et al. (2002, in prep.). See also Veilleux
(2001).}
\end{figure}

\subsection{Metallicity Effects}

The line ratio diagnostics discussed in this review often are a
sensitive function of the metal contents in the ionized gas (see,
e.g., Ferland \& Netzer 1983; Veilleux \& Osterbrock 1987 for early
papers describing the effects of metallicity).  The metallicity is
well known to be correlated positively with the mass of the host
galaxies (e.g., Bender, Burnstein, \& Faber 1993), although this
result has only been proven at low redshifts. In the early universe,
one would expect declining metal abundances with increasing
redshifts. The redshift dependence of the relative abundances of the
elements involved in the emission-line ratios is a complex function of
the star formation history and chemical evolution (including the effects of
gas accretion and outflows) of the host galaxy environment (see, e.g.,
Hamann \& Ferland 1999 for a discussion of QSO hosts). The usefulness
of emission-line diagnostics at high redshifts will directly depend on
the availability of accurate metallicity measurements and diagnostic
tools properly calibrated in terms of metallicity.

\section{Summary}

UV--Optical--IR emission-line ratios are powerful diagnostics tools to
discriminate between starbursts and AGNs. The following ratios have been
shown to be the most reliable tools for this purpose.
\begin{itemize}
\item[1.] Ultraviolet: N V $\lambda$1240/Ly$\alpha$, N V
$\lambda$1240/He II $\lambda$1640, C IV $\lambda$1548/Ly$\alpha$.
\item[2.] Optical: [O III] $\lambda$5007//H$\beta$, [N II]
$\lambda$6583/H$\alpha$, [S II] $\lambda\lambda$6724/H$\alpha$, [O I]
$\lambda$6300/H$\alpha$, [O II] $\lambda\lambda$7324/H$\alpha$, [Fe
VII] $\lambda$6087/H$\alpha$, [Ne V] $\lambda$3426/H$\beta$, He II
$\lambda$4686/H$\beta$.
\item[3.] Near-infrared: Obscured broad Pa$\alpha$ 1.875 $\mu$m, [Si
VI] 1.962 $\mu$m/Pa$\alpha$.
\item[4.] Mid-Infrared: [Ne V] 14 $\mu$m/[Ne II] 12.8 $\mu$m, [O IV]
26 $\mu$m/[Ne II] 12.8 $\mu$m, EW(PAH 7.7 $\mu$m), overall SEDs
especially 25 $\mu$m/60 $\mu$m colors.
\end{itemize}
A number of issues complicate the use of line ratios as discriminants
between starburst and active galaxies, but additional measures can be
used to clarify the situation:
\begin{itemize}
\item[1.] Shock ionization: If shocks are important, one would
generally expect correlations between the line ratios and gas
kinematics, a UV continuum extended on the same scale as the shock
structure, and high gas temperatures.
\item[2.] Aperture effects: One should use a constant linear aperture
to avoid variations in the contributions from circumnuclear
starbursts.
\item[3.] Morphological bias: The spectral classification is likely to
depend on the morphology of the host, especially the merger
phase. Selection methods based on morphology will bias the sample.
\item[4.] Metallicity: Massive host galaxies in the local universe
have larger metallicity, but high-redshift galaxies should be less
dusty and less metal rich. One needs to use emission-line diagnostics
which are properly calibrated as a function of metallicity and
reddening.
\end{itemize}

Several new instruments will help refine the diagnostic tools
discussed in this paper. The Cosmic Origins Spectrograph (COS), to be
installed in 2003 on HST, will provide the high ultraviolet throughput
needed to calibrate the UV diagnostic tools as a function of
metallicity, evaluate the importance of shock ionization with the use
of the C III and NIII temperature-sensitive line ratios, and to help
resolve the circumnuclear starbursts and shock-excited winds around
AGNs.  The advent of SIRTF will help in the calibration of the
infrared diagnostic tools as a function of metallicity {\em and dust
extinction}. This spacecraft will also be a powerful instrument to
search for infrared-bright AGNs. Ground-based work with adaptive
optics and integral-field units will improve the sensitivity of
searches for obscured AGNs by focussing on the inner regions of
galaxies and avoiding the circumnuclear material associated with other
phenomena. Spectroscopic follow-ups from the ground will help identify
and classify AGN candidates in space-based and submm-selected samples. 

\begin{acknowledgments}
The ground-based study on ultraluminous infrared galaxies discussed in
this paper is done in collaboration with Drs.~D. B. Sanders and
D.-C. Kim.  The author gratefully acknowledges the financial support
of NASA through LTSA grant number NAG~56547.
\end{acknowledgments}

\end{document}